\documentclass[12pt,preprint]{aastex}

\usepackage{emulateapj5}           

\slugcomment{Accepted for Astrophysical Journal Letters, draft version from \today}

\shorttitle{C$_2$H in massive star formation}
\shortauthors{Beuther et al.}

\begin{document}

\title{Ethynyl (C$_2$H) in massive star formation: Tracing the initial conditions?}


\author{H.~Beuther, D. Semenov, Th. Henning, H. Linz}
\affil{Max Planck Institute for Astronomy, K\"onigstuhl 17, 69117 Heidelberg, Germany}

\begin{abstract}
  APEX single-dish observations at sub-millimeter wavelengths toward a
  sample of massive star-forming regions reveal that C$_2$H is almost
  omni-present toward all covered evolutionary stages from Infrared
  Dark Clouds via High-Mass Protostellar Objects to Ultracompact H{\sc
    ii} regions. High-resolution data from the Submillimeter Array
  toward one hot-core like High-Mass Protostellar Object show a
  shell-like distribution of C$_2$H with a radius of $\sim$9000\,AU
  around the central submm peak position.  These observed features are
  well reproduced by a 1D cloud model with power-law density and
  temperature distributions and a gas-grain chemical network. The
  reactive C$_2$H radical (ethynyl) is abundant from the onset of
  massive star formation, but later it is rapidly transformed to other
  molecules in the core center. In the outer cloud regions the
  abundance of C$_2$H remains high due to constant replenishment of
  elemental carbon from CO being dissociated by the interstellar UV
  photons. We suggest that C$_2$H may be a molecule well suited to
  study the initial conditions of massive star formation.
\end{abstract}

\keywords{stars: formation -- astrochemistry -- stars: individual (IRAS18089-1732)}

\section{Introduction}

Spectral line surveys have revealed that high-mass star-forming
regions are rich reservoirs of molecules from simple diatomic species
to complex and larger molecules (e.g.,
\citealt{schilke1997b,hatchell1998b,comito2005,bisschop2007}).
However, there have been rarely studies undertaken to investigate the
chemical evolution during massive star formation from the earliest
evolutionary stages, i.e., from High-Mass Starless Cores (HMSCs) and
High-Mass Cores with embedded low- to intermediate-mass protostars
destined to become massive stars, via High-Mass Protostellar Objects
(HMPOs) to the final stars that are able to produce Ultracompact H{\sc
  ii} regions (UCH{\sc ii}s, see \citealt{beuther2006b} for a recent
description of the evolutionary sequence). The first two evolutionary
stages are found within so-called Infrared Dark Clouds (IRDCs).  While
for low-mass stars the chemical evolution from early molecular
freeze-out to more evolved protostellar cores is well studied (e.g.,
\citealt{bergin1997,dutrey1997,pavlyuchenkov2006,joergensen2007}),
it is far from clear whether similar evolutionary patterns are present
during massive star formation.

To better understand the chemical evolution of high-mass star-forming
regions we initiated a program to investigate the chemical properties
from IRDCs to UCH{\sc ii}s from an observational and theoretical
perspective. We start with single-dish line surveys toward a large
sample obtaining their basic characteristics, and then perform
detailed studies of selected sources using interferometers on smaller
scales. These observations are accompanied by theoretical modeling of
the chemical processes.  Long-term goals are the chemical
characterization of the evolutionary sequence in massive star
formation, the development of chemical clocks, and the identification
of molecules as astrophysical tools to study the physical processes
during different evolutionary stages. Here, we present an initial
study of the reactive radical ethynyl (C$_2$H) combining single-dish
and interferometer observations with chemical modeling.  Although
C$_2$H was previously observed in low-mass cores and Photon Dominated
Regions (e.g., \citealt{millar1984,jansen1995}), so far it was not
systematically investigated in the framework of high-mass star
formation.

\section{Observations}
\label{obs}

The 21 massive star-forming regions were observed with the Atacama
Pathfinder Experiment (APEX) in the 875\,$\mu$m window in fall 2006.
We observed 1\,GHz from 338 to 339\,GHz and 1\,GHz in the image
sideband from 349 to 350\,GHz.  The spectral resolution was
0.1\,km\,s$^{-1}$, but we smoothed the data to
$\sim$0.9\,km\,s$^{-1}$. The average system temperatures were around
200\,K, each source had on-source integration times between 5 and 16
min. The data were converted to main-beam temperatures with forward
and beam efficiencies of 0.97 and 0.73, respectively
\citep{belloche2006}. The average $1\sigma$ rms was 0.4\,K.  The main
spectral features of interest are the C$_2$H lines around 349.4\,GHz
with upper level excitation energies $E_u/k$ of 42\,K (line blends of
C$_2$H$(4_{5,5}-3_{4,4})$ \& C$_2$H$(4_{5,4}-3_{4,3})$ at
349.338\,GHz, and C$_2$H$(4_{4,4}-3_{3,3})$ \&
C$_2$H$(4_{4,3}-3_{3,2})$ at 349.399\,GHz). The beam size was $\sim
18''$.

The original Submillimeter Array (SMA) C$_2$H data toward the
HMPO\,18089-1732 were first presented in \citet{beuther2005c}. There
we used the compact and extended configurations resulting in good
images for all spectral lines except of C$_2$H. For this project, we
re-worked on these data only using the compact configuration. Because
the C$_2$H emission is distributed on larger scales (see
\S\ref{results}), we were now able to derive a C$_2$H image. The
integration range was from 32 to 35\,km\,s$^{-1}$, and the achieved
$1\sigma$ rms of the C$_2$H image was 450\,mJy\,beam$^{-1}$.  For more
details on these observations see \citet{beuther2005c}.

\section{Results}
\label{results}

The sources were selected to cover all evolutionary stages from IRDCs
via HMPOs to UCH{\sc ii}s. We derived our target list from the samples
of \citet{klein2005,fontani2005,hill2005,beltran2006}.  Table
\ref{sample} lists the observed sources, their coordinates, distances,
luminosities and a first order classification into the evolutionary
sub-groups IRDCs, HMPOs and UCH{\sc ii}s based on the previously
available data. Although this classification is only based on a
limited set of data, here we are just interested in general
evolutionary trends. Hence, the division into the three main classes
is sufficient.

\begin{figure}
\includegraphics[angle=-90,width=8.8cm]{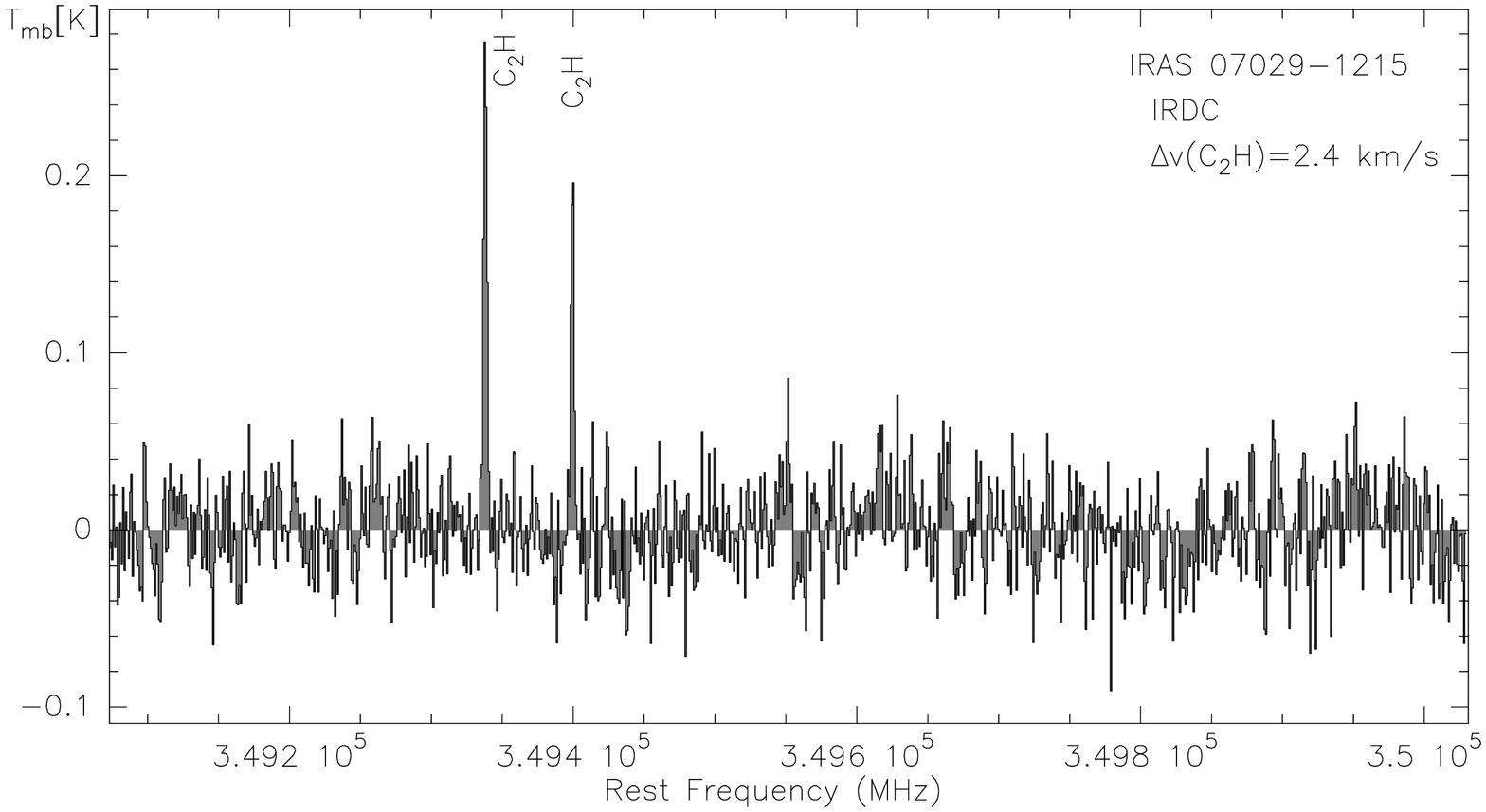}\\
\includegraphics[angle=-90,width=8.8cm]{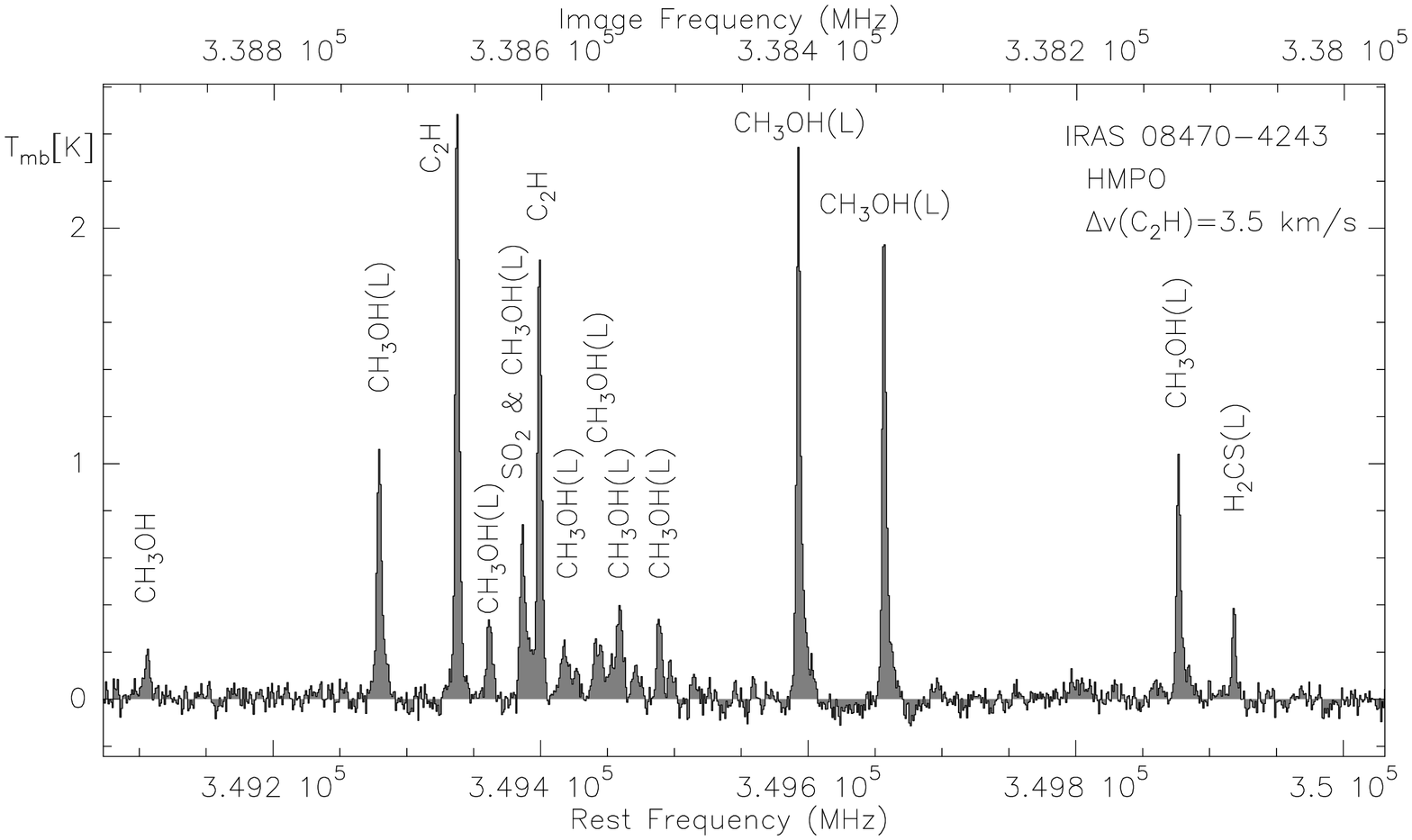}\\
\includegraphics[angle=-90,width=8.8cm]{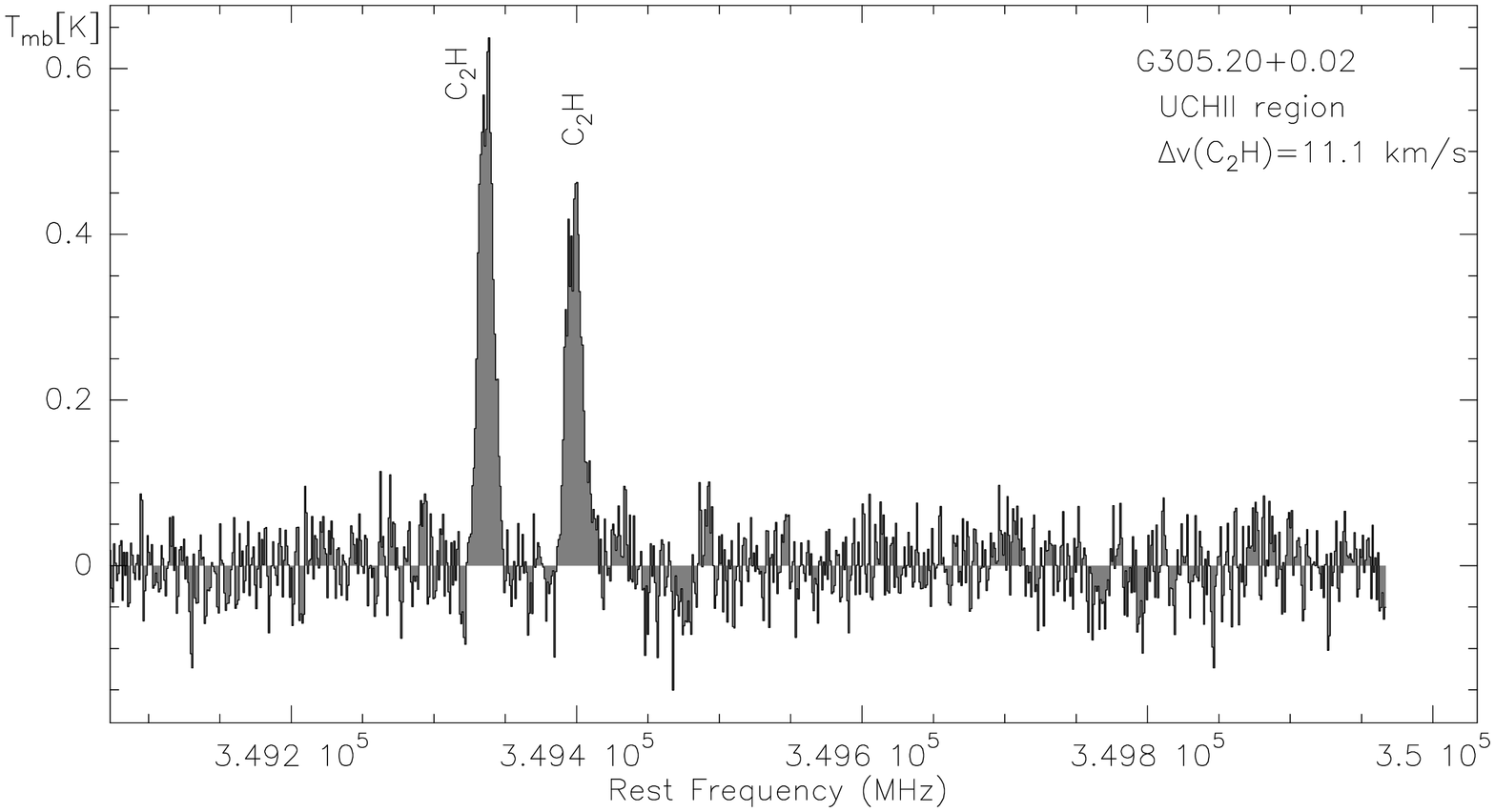}
\caption{Sample spectra obtained with APEX in a double-sideband mode.
  The spectra cover 1\,GHz of data around 349.5\,GHz and 1\,GHz around
  338.5\,GHz. All spectral lines are labeled, lines in the lower
  sideband are marked with a ``(L)''. The top-panel shows an IRDC
  example, the middle-panel a typical HMPO/hot core and the bottom
  panel a UCH{\sc ii} region. The C$_2$H line-widths are indicated in
  each panel.}
\label{spectra}
\end{figure}

Figure \ref{spectra} presents sample spectra toward one source of each
evolutionary group. While we see several CH$_3$OH lines as well as
SO$_2$ and H$_2$CS toward some of the HMPOs and UCH{\sc ii}s but not
toward the IRDCs, the surprising result of this comparison is the
presence of the C$_2$H lines around 349.4\,GHz toward all source types
from young IRDCs via the HMPOs to evolved UCH{\sc ii}s.  Table
\ref{sample} lists the peak brightness temperatures, the integrated
intensities and the FWHM line-widths of the C$_2$H line blend at
349.399\,GHz. The separation of the two lines of 1.375\,MHz already
corresponds to a line-width of 1.2\,km\,s$^{-1}$. We have three C$_2$H
non-detections (2 IRDCs and 1 HMPO), however, with no clear trend with
respect to the distances or the luminosities (the latter comparison is
only possible for the HMPOs). While IRDCs are on average colder than
more evolved sources, and have lower brightness temperatures, the
non-detections are more probable due to the relatively low sensitivity
of the short observations (\S\ref{obs}). Hence, the data indicate
that the C$_2$H lines are detected independent of the evolutionary
stage of the sources in contrast to the situation with other
molecules.  When comparing the line-widths between the different
sub-groups, one finds only a marginal difference between the IRDCs and
the HMPOs (the average $\Delta v$ of the two groups are 2.8 and
3.1\,km\,s$^{-1}$).  However, the UCH{\sc ii}s exhibit significantly
broader line-widths with an average value of 5.5\,km\,s$^{-1}$.

Intrigued by this finding, we wanted to understand the C$_2$H spatial
structure during the different evolutionary stages.  Therefore, we
went back to a dataset obtained with the Submillimeter Array toward
the hypercompact H{\sc ii} region IRAS\,18089-1732 with a much higher
spatial resolution of $\sim 1''$ \citep{beuther2005c}.  Albeit this
hypercompact H{\sc ii} region belongs to the class of HMPOs, it is
already in a relatively evolved stage and has formed a hot core with a
rich molecular spectrum.  \citet{beuther2005c} showed the spectral
detection of the C$_2$H lines toward this source, but they did not
present any spatially resolved images. To recover large-scale
structure, we restricted the data to those from the compact SMA
configuration (\S\ref{obs}). With this refinement, we were able to
produce a spatially resolved C$_2$H map of the line blend at
349.338\,GHz with an angular resolution of $2.9''\times 1.4''$
(corresponding to an average linear resolution of 7700\,AU at the
given distance of 3.6\,kpc). Figure \ref{18089} presents the
integrated C$_2$H emission with a contour overlay of the 860\,$\mu$m
continuum source outlining the position of the massive protostar. In
contrast to almost all other molecular lines that peak along with the
dust continuum \citep{beuther2005c}, the C$_2$H emission surrounds the
continuum peak in a shell-like fashion.

\section{Discussion and Conclusions}

To understand the observations, we conducted a simple chemical
modeling of massive star-forming regions. A 1D cloud model with a mass
of 1200\,M$_\sun$, an outer radius of 0.36\,pc and a power-law density
profile ($\rho\propto r^p$ with $p=-1.5$) is the initially assumed
configuration. Three cases are studied: (1) a cold isothermal cloud
with $T=10$\,K, (2) $T=50$\,K, and (3) a warm model with a temperature
profile $T\propto r^q$ with $q=-0.4$ and a temperature at the outer
radius of 44\,K. The cloud is illuminated by the interstellar UV
radiation field (IRSF, \citealt{draine1978}) and by cosmic ray
particles (CRP). The ISRF attenuation by single-sized $0.1\mu$m
silicate grains at a given radius is calculated in a plane-parallel
geometry following \citet{vandishoeck1988}. The CRP ionization rate is
assumed to be $1.3\times 10^{-17}$~s$^{-1}$ \citep{spitzer1968}. The
gas-grain chemical model by \citet{vasyunin2008} with the desorption
energies and surface reactions from \citet{garrod2006} is used.
Gas-phase reaction rates are taken from RATE\,06 \citep{woodall2007},
initial abundances, were adopted from the ``low metal'' set of
\citet{lee1998}.

\begin{figure}
\includegraphics[width=8.8cm]{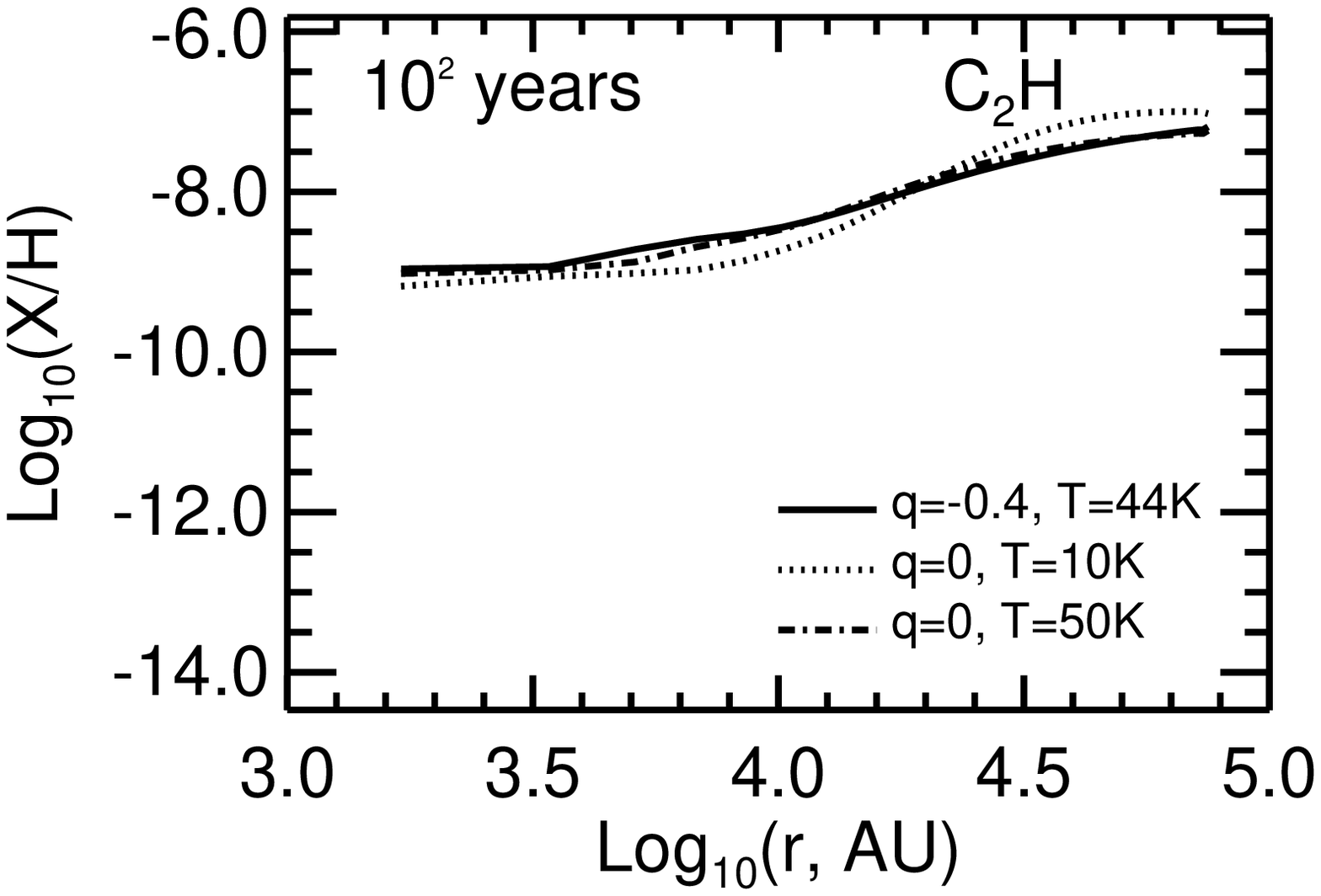}\\
\includegraphics[width=8.8cm]{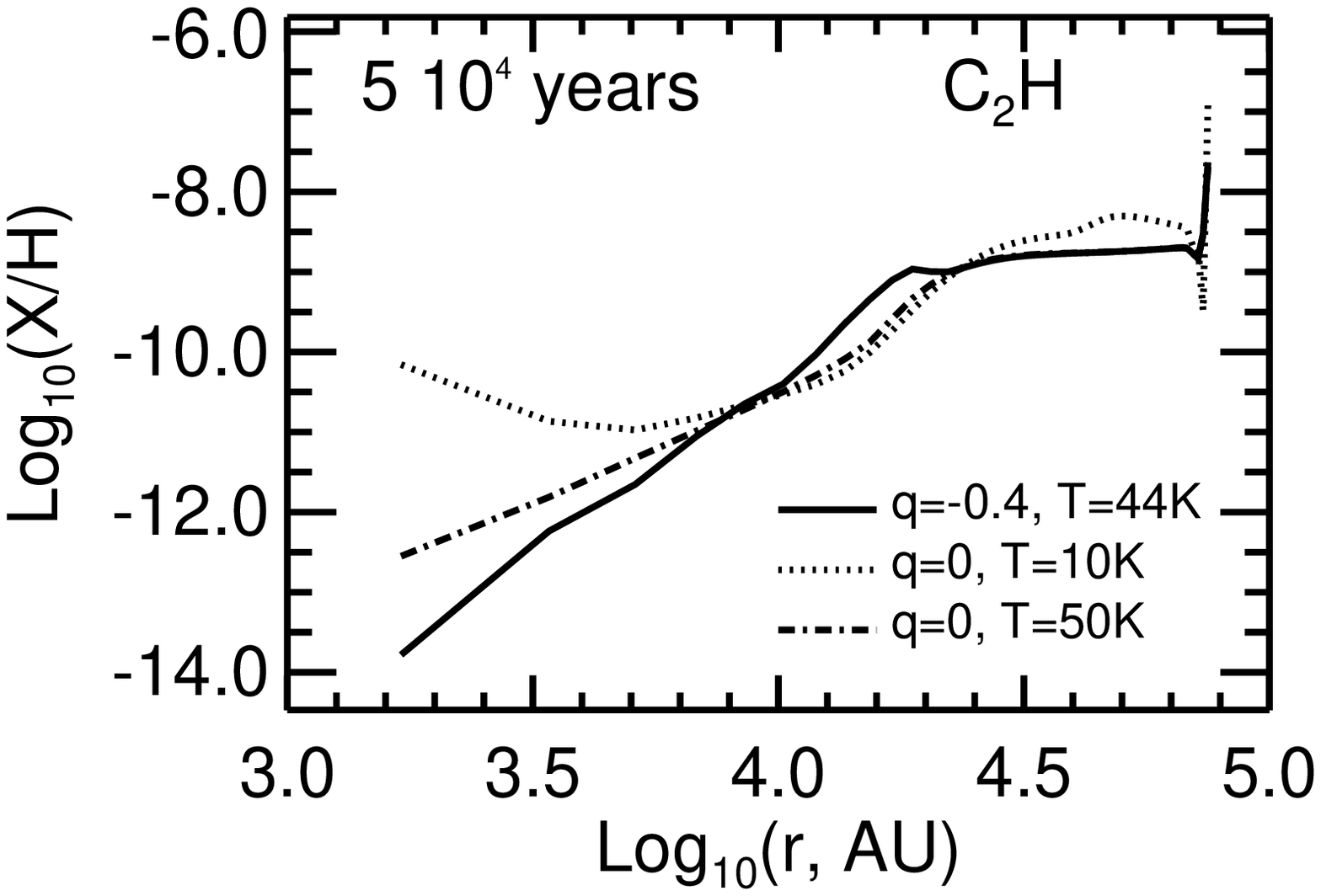}\\
\caption{Top left: Radial profiles of the C$_2$H abundance for the
  three cloud models at early times, $t \la 100$~yr. Bottom left: The
  same but for the later time step of $5\times 10^4$\,yr. The parameter $q$
  denotes the exponent of the assumed temperature distribution (see
  also main text).
}
\label{model}
\end{figure}

Figure \ref{model} presents the C$_2$H abundances for the three models
at two different time steps: (a) 100\,yr, and (b) in a more evolved
stage after $5\times10^4$\,yr. The C$_2$H abundance is high toward the
core center right from the beginning of the evolution, similar to
previous models (e.g., \citealt{millar1985,herbst1986,turner1999}).
During the evolution, the C$_2$H abundance stays approximately
constant at the outer core edges, whereas it decreases by more than
three orders of magnitude in the center, except for the cold $T=10$~K
model.  The C$_2$H abundance profiles for all three models show
similar behavior.

The chemical evolution of ethynyl is determined by relative removal
rates of carbon and oxygen atoms or ions into molecules like CO, OH,
H$_2$O. Light ionized hydrocarbons CH$^+_{\rm n}$ (n=2..5) are quickly
formed by radiative association of C$^+$ with H$_2$ and hydrogen
addition reactions: C$^+$ $\rightarrow$ CH$_2^+$ $\rightarrow$
CH$_3^+$ $\rightarrow$ CH$_5^+$.  The protonated methane reacts with
electrons, CO, C, OH, and more complex species at later stage and
forms methane.  The CH$_4$ molecules undergo reactive collisions with
C$^+$, producing C$_2$H$_2^+$ and C$_2$H$_3^+$. An alternative way to
produce C$_2$H$_2^+$ is the dissociative recombination of CH$_5^+$
into CH$_3$ followed by reactions with C$^+$.  Finally, C$_2$H$_2^+$
and C$_2$H$_3^+$ dissociatively recombine into CH, C$_2$H, and
C$_2$H$_2$. The major removal for C$_2$H is either the direct
neutral-neutral reaction with O that forms CO, or the same reaction
but with heavier carbon chain ions that are formed from C$_2$H by
subsequent insertion of carbon. At later times, depletion and
gas-phase reactions with more complex species may enter into this
cycle.  At the cloud edge the interstellar UV radiation
instantaneously dissociates CO despite its self-shielding,
re-enriching the gas with elemental carbon.

The transformation of C$_2$H into CO and other species proceeds
efficiently in dense regions, in particular in the ``warm'' model
where endothermic reactions result in rich molecular complexity of the
gas (see Fig.~\ref{model}).  In contrast, in the ``cold'' 10\,K model
gas-grain interactions and surface reactions become important.  As a
result, a large fraction of oxygen is locked in water ice that is hard
to desorb ($E_{\rm des} \sim 5500$~K), while half of the elemental
carbon goes to volatile methane ice ($E_{\rm des} \sim 1300$~K). Upon
CRP heating of dust grains, this leads to much higher gas-phase
abundance of C$_2$H in the cloud core for the cold model compared to
the warm model. The effect is not that strong for less dense regions
at larger radii from the center.

Since the C$_2$H emission is anti-correlated with the dust continuum
emission in the case of IRAS\,18089-1732 (Fig.\,\ref{18089}), we do
not have the H$_2$ column densities to quantitatively compare the
abundance profiles of IRAS\,18089-1732 with our model. However, data
and model allow a qualitative comparison of the spatial structures.
Estimating an exact evolutionary time for IRAS\,18089-1732 is hardly
possible, but based on the strong molecular line emission, its high
central gas temperatures and the observed outflow-disk system
\citep{beuther2004a,beuther2004b,beuther2005c}, an approximate age of
$5\times10^4$\,yr appears reasonable.  Although dynamical and chemical
times are not necessarily exactly the same, in high-mass star
formation they should not differ to much: Following the models by
\citet{mckee2003} or \citet{krumholz2006b}, the luminosity rises
strongly right from the onset of collapse which can be considered as a
starting point for the chemical evolution. At the same time disks and
outflows evolve, which should hence have similar time-scales.  The
diameter of the shell-like C$_2$H structure in IRAS\,18089-1732 is
$\sim 5''$ (Fig.\,\ref{18089}), or $\sim$9000\,AU in radius at the
given distance of 3.6\,kpc.  This value is well matched by the modeled
region with decreased C$_2$H abundance (Fig.\,\ref{model}).  Although
in principle optical depths and/or excitation effects could mimic the
C$_2$H morphology, we consider this as unlikely because the other
observed molecules with many different transitions all peak toward the
central submm continuum emission in IRAS\,18089-1732
\citep{beuther2005c}. Since C$_2$H is the only exception in that rich
dataset, chemical effects appear the more plausible explanation.

The fact that we see C$_2$H at the earliest and the later evolutionary
stages can be explained by the reactive nature of C$_2$H: it is
produced quickly early on and gets replenished at the core edges by
the UV photodissociation of CO. The inner ``chemical'' hole observed
toward IRAS\,18089-1732 can be explained by C$_2$H being consumed in
the chemical network forming CO and more complex molecules like larger
carbon-hydrogen complexes and/or depletion.

The data show that C$_2$H is not suited to investigate the central gas
cores in more evolved sources, however, our analysis indicates that
C$_2$H may be a suitable tracer of the earliest stages of (massive)
star formation, like N$_2$H$^+$ or NH$_3$ (e.g.,
\citealt{bergin2002,tafalla2004,beuther2005a,pillai2006}). While a
spatial analysis of the line emission will give insights into the
kinematics of the gas and also the evolutionary stage from chemical
models, multiple C$_2$H lines will even allow a temperature
characterization. With its lowest $J=1-0$ transitions around 87\,GHz,
C$_2$H has easily accessible spectral lines in several bands between
the 3\,mm and 850\,$\mu$m.  Furthermore, even the 349\,GHz lines
presented here have still relatively low upper level excitation
energies ($E_u/k\sim42$\,K), hence allowing to study cold cores even
at sub-millimeter wavelengths.  This prediction can further be proved
via high spectral and spatial resolution observations of different
C$_2$H lines toward young IRDCs.

\acknowledgments{H.B. acknowledges financial support
  by the Emmy-Noether-Programm of the Deutsche Forschungsgemeinschaft
  (DFG, grant BE2578). }


\clearpage

\begin{table}[htb]
{\small
\begin{center}
\caption{Source parameters}
\begin{tabular}{lrrrrrrrrr}
\hline
\hline
Name & Type & R.A.    & Dec.    & $d$   & $L$           & $T_{\rm{mb}}$ & $\int T_{\rm{mb}}dv$ & $\Delta v(\rm{C_2H})$ & Ref.\\
     &      & J2000 & J2000 & kpc & log(L$_{\odot}$) &    K        & K\,km\,s$^{-1}$    &  km\,s$^{-1}$       & \\
\hline
IRAS07029 & IRDC        & 07:05:11.1 & -12:19:02 & 1.0         & $^b$ & 0.21 & 0.52 & $2.4\pm 0.3$ & (1) \\
IRAS08477 & IRDC        & 08:49:32.9 & -44:10:47 & 1.8         & $^b$ & 0.23 & 0.76 & $3.1\pm 0.5$ & (2,3) \\
IRAS09014 & IRDC        & 09:03:09.8 & -47:48:28 & 1.3         & $^b$ & --   & --   & --     & (2,3)\\
IRAS13039 & IRDC        & 13:07:07.0 & -61:24:47 & 2.4         & $^b$ & --   & --   & --     & (2,3)\\
IRAS14000 & IRDC        & 14:03:36.6 & -61:18:28 & 5.6         & $^b$ & 0.38 & 1.13 & $2.8\pm 0.5$ & (2,3)\\
IRAS08211 & HMPO        & 08:22:52.3 & -42:07:57 & 1.7         & 3.5  & 0.38 & 0.85 & $2.1\pm 0.2$ & (2,3)\\
IRAS08470 & HMPO        & 08:48:47.9 & -42:54:22 & 2.2         & 4.2  & 1.85 & 6.81 & $3.5\pm 0.4$ & (3)\\
IRAS08563 & HMPO        & 08:58:12.5 & -42:37:34 & 1.7         & 3.2  & 1.08 & 3.30 & $2.9\pm 0.1$ & (2,3)\\
IRAS09131 & HMPO        & 09:14:55.5 & -47:36:13 & 1.7         & 3.4  & 0.23 & 0.50 & $2.1\pm 0.3$ & (2,3)\\
IRAS09209 & HMPO        & 09:22:34.6 & -51:56:26 & 6.4         & 4.1  & --   & --   & --           & (2,3)\\
IRAS09578 & HMPO        & 09:59:31.0 & -57:03:45 & 1.7         & 3.9  & 0.21 & 0.64 & $2.8\pm 0.4$ & (3)\\
IRAS10123 & HMPO        & 10:14:08.8 & -57:42:12 & 0.9/3.0$^c$ & 3.4/4.4$^c$ & 0.17$^a$ & 0.46 & $2.5\pm 0.4$ & (2,3)\\
IRAS10184 & HMPO        & 10:20:14.7 & -58:03:38 & 5.4         & 5.5  & 0.41 & 1.68 & $3.9\pm 0.3$ & (3)\\
IRAS10276 & HMPO        & 10:29:30.1 & -57:26:40 & 5.9         & 4.9  & 0.24$^a$ & 0.67 & $2.6\pm 0.5$ & (3)\\
IRAS10295 & HMPO        & 10:31:28.3 & -58:02:07 & 5.0         & 5.8  & 0.69 & 3.45 & $4.7\pm 0.3$ & (3)\\
IRAS10320 & HMPO        & 10:33:56.4 & -59:43:53 & 9.1         & 5.4  & 0.85 & 3.48 & $3.8\pm 0.6$ & (3)\\
G294.97   & UCH{\sc ii} & 11:39:09.0 & -63:28:38 & 1.3/5.8$^c$ & 3.9/5.3$^c$ & 0.31 & 0.78 & $2.3\pm 0.4$ & (4)\\
G305.20   & UCH{\sc ii} & 13:11:12.3 & -62:44:57 & 3.0/6.8$^c$ & 5.1/6.1$^c$ & 0.44 & 5.21 & $11.1\pm 0.5$& (4)\\
G305.37   & UCH{\sc ii} & 13:12:36.3 & -62:33:39 & 3.0/6.8$^c$ & $^d$        & 1.24 & 6.28 & $4.8\pm 0.2$ & (4)\\
G305.561  & UCH{\sc ii} & 13:14:25.8 & -62:44:32 & 4.0         & 5.1  & 0.94 & 4.71 & $4.7\pm 0.6$ & (4,5)\\
IRAS14416 & UCH{\sc ii} & 14:45:22.0 & -59:49:39 & 2.8         & 5.1  & 1.24 & 6.19 & $4.7\pm 0.6$ & (6)\\
\hline
\hline
\label{sample}
\end{tabular}
\end{center}
~\\Ref.: (1) \citet{klein2005}, (2) \citet{fontani2005}, (3) \citet{beltran2006}, (4) \citet{hill2005}, (5) \citet{faundez2004}, (6) \citet{vig2007}.\\
$^a$ For these sources we list the parameters of the C$_2$H line
blend at 349.338\,GHz, for all other sources it is the C$_2$H line at
349.399\,GHz.\\
$^b$ Since the IRDCs are per default not detected at short wavelengths, they are no IRAS source and we cannot derive a luminosity.\\
$^c$ Near and far distances and corresponding luminosities.\\
$^d$ No IRAS counterpart, hence no luminosity estimate.
}
\end{table}

\clearpage

\begin{figure}
\caption{The grey-scale shows the integrated emission (from 32 to
  35\,km\,s$^{-1}$) of the line blend of C$_2$H$(4_{5,5}-3_{4,4})$ and
  C$_2$H$(4_{5,4}-3_{4,3})$ around 349.338\,GHz obtained with the
  Submillimeter Array using only the compact configuration data
  \citep{beuther2005c}. The resulting synthesized beam is shown at the
  bottom left corner ($2.9''\times 1.4''$). The C$_2$H emission is
  presented in grey-scale with thin contours (dashed contours negative
  features), and the 860\,$\mu$m continuum peak is shown in thick
  contours. The C$_2$H emission starts at the $2\sigma$ level and
  continues in $1\sigma$ steps with the $1\sigma$ level of
  450\,mJy\,beam$^{-1}$.  The 860\,$\mu$m emission is contoured from
  10 to 90\% (step 10\%) of the peak emission of
  1.4\,Jy\,beam$^{-1}$.}
\label{18089}
\end{figure}

\end{document}